\documentclass[superscriptaddress, reprint, amsmath, amssymb, aps, pra, floatfix]{revtex4-2}

\usepackage[colorlinks=true]{hyperref}
\usepackage{graphicx}
\usepackage{dcolumn}
\usepackage{bm}
\usepackage{orcidlink}
\usepackage{physics}
\usepackage{color}
\usepackage{amsthm}
\usepackage{multirow}
\newtheorem{thm}{Theorem}

\begin{document}


\title{Spectral Gaps via Imaginary Time}

\author{Jacob M. Leamer\orcidlink{0000-0002-2125-7636}}
    \email{leamerjm@gmail.com}
\affiliation{Advanced Photonic and Electronic Sciences Division, US DEVCOM Army Research Laboratory, Adelphi, MD 20783, USA}
\author{Alicia B. Magann\orcidlink{0000-0002-1402-3487}}%
\affiliation{Quantum Algorithms and Applications Collaboratory, Sandia National Laboratories, Albuquerque, New Mexico 87185, USA}%


\author{Gerard McCaul\orcidlink{0000-0001-7972-456X}}
\affiliation{Department of Physics, Loughborough University, Loughborough, UK}

\author{Denys I. Bondar\orcidlink{0000-0002-3626-4804}}%
    \email{dbondar@tulane.edu}
\affiliation{Department of Physics and Engineering Physics, Tulane University, 6823 St. Charles Ave., New Orleans, LA 70118, USA}

\date{\today}

\begin{abstract}
The spectral gap occupies a role of central importance in many open problems in physics. We present an approach for evaluating the spectral gap of a Hamiltonian from a simple ratio of two expectation values, both of which are evaluated using a quantum state that is evolved in imaginary time. In principle, the only requirement is that the initial state is supported on both the ground and first excited states.  We demonstrate this approach for the Fermi-Hubbard and transverse-field Ising models through numerical simulation. We then go on to explore avenues for its implementation on quantum computers using imaginary-time quantum dynamical emulation.

\end{abstract}

\maketitle
\section{\label{sec:intro}Introduction}

Excited states play a pivotal role in many applications. Very often, however, the physical processes of interest in applications depend not on the absolute energy of the excited state but instead, on the \textit{gap} between this energy and the ground state energy. Explicitly, the determining feature for such systems is the spectral gap
\begin{align}
	\Delta E = E_1 - E_0,
\end{align}
where $E_0$ and $E_1$ are the energies of the ground and first excited states, respectively. 

In molecular and materials contexts, these lowest two eigenvalues of a Hamiltonian $H$ set fluorescence  and phosphorescence wavelengths \cite{Williams371} and excitonic energy-transfer windows \cite{jacquemin_excited-state_2011}, as well as transport and optical properties in correlated-electron systems and Mott insulators \cite{imada_1998,anderson_1987,hybertsen_first-principles_1985,hybertsen_electron_1986,godby_accurate_1986,godby_self-energy_1988}. In lattice models, the gap fixes correlation lengths and phase structure, and so plays a central role in identifying and characterising exotic phases such as topological insulators, spin liquids, and other frustrated magnets \cite{yang_topological_2013,hastings_spectral_2006,balents_spin_2010,han_fractionalized_2012,hofrichter_2016}. In quantum information settings, the gap sets adiabatic run times, controls diabatic errors in quantum annealing, and determines state-transfer fidelities along spin chains and related architectures \cite{hartmann_excitation_2006,farhi_quantum_2000,troels_2014,shin_how_2014,boixo_evidence_2014}.

From a computational point of view, the same quantity is often the bottleneck. In the worst case, deciding whether a local Hamiltonian is gapped or gapless can be undecidable \cite{cubitt2015undecidability}, while estimating low-lying spectra of generic many-body systems is encapsulated by the local Hamiltonian problem. This is a canonically computationally `hard' task \cite{Kempe2006LocalHam}, and even discounting questions of undecidability, answering this with standard numerical tools involves stringent tradeoffs. Fermionic quantum Monte Carlo suffers from the sign problem, which is NP-hard to cure in general \cite{Troyer2005SignProblem}. Tensor-network methods, including matrix product states and projected entangled pair states, tame the Hilbert-space dimension but pay directly in bond dimension and entanglement structure \cite{Orus2014TN, lukin_spectral_2024, bilokon_dispersion_2025}. A variety of techniques have been developed to resolve spectra, ranging from Green's function \cite{hybertsen_first-principles_1985,hybertsen_electron_1986,godby_accurate_1986,godby_self-energy_1988}, equation-of-motion and configuration-interaction schemes \cite{jacquemin_excited-state_2011,ramos_low-lying_2018,ferre_density-functional_2016,kaldor_degenerate_1975,blume_excited_1997,luchow_computing_2003}, to a burgeoning zoo of quantum algorithms \cite{hunt_quantum_2018,motta_determining_2020,kamakari_digital_2022,turro_imaginary-time_2022}. What unites these approaches is that they must all pay a price for the complexity of the task they target, but are taxed in a currency peculiar to each, paying in system size, basis size, coherence time, or measurement overhead. 

Within this family of techniques, imaginary time evolution may be considered one of the oldest, and most robust \cite{feynman_hibbs_1965}. It has been used extensively in path–integral \cite{negele_orland_1988,ceperley_1995,foulkes_2001} and influence functional formulations \cite{mccaul_partition-free_2017,mccaul_how_2021}, where a Wick rotation maps $t \mapsto -i\tau$. The underlying Schr\"odinger equation becomes a diffusion equation, and higher-energy eigenstates are exponentially suppressed as a function of $\tau$. On classical hardware, this underpins a wide range of ground-state and low-energy solvers for continuum and lattice systems \cite{chin_any_2009,lehtovaara_solution_2007}, as well as the imaginary-time components of many quantum Monte Carlo algorithms. On quantum hardware, the last few years have seen a parallel development of quantum imaginary time evolution (QITE) and related flows \cite{motta_determining_2020,kamakari_digital_2022,turro_imaginary-time_2022,McMahon2025QITEFlows,Alipour2025StateBasedITE}, which trade exact unitarity for effective Euclidean-time dynamics implemented via local updates or linear combinations of unitaries. These methods are attractive precisely because they convert spectral questions into questions about relaxation along a one-parameter flow.

In parallel, there is an emerging picture in which ``effective Hamiltonians'' are used less as fundamental generators and more as workhorses for solving differential equations. Block-encoding and linear-combination-of-unitaries (LCU) constructions have been used to map large classes of linear differential equations - including Carleman-linearised nonlinear systems - into Hamiltonian simulation tasks amenable to near-future quantum algorithms \cite{Wu2025EffectiveHamiltonians}. A complementary route is based on Koopman–von Neumann and Liouville embeddings, constructing quantum or quantum-inspired operators whose spectra encode transport, advection–diffusion, and fusion-relevant dynamics \cite{Joseph2020KvN,NovikauJoseph2025AdvecDiff,NovikauJoseph2025Carleman,Joseph2023Fusion,ParkerJoseph2020GenEigQPE}. In both lines of work, ``solving the PDE'' becomes a spectral problem for an auxiliary generator; what matters operationally are low-lying eigenvalues, gaps, and resolvents, not the full many-body spectrum of a physical Hamiltonian.

In recent work \cite{McCaul2025FreeSnacks}, the Imaginary-Time Quantum Dynamical Emulation (ITQDE) \cite{leamer_quantum_2024} method was analysed as a general framework for reconstructing spectral information. ITQDE is a non-variational, sampling-based technique, in which a characteristic ``spectral staircase'' is constructed from the overlap of oppositely propagated states. There it was shown that while one cannot cheat the worst-case complexity of spectral estimation - there is no free lunch - there is nevertheless a \textit{free snack}: controllable approximations to spectral quantities that would otherwise appear to demand the resolution of individual eigenstates via quantum phase estimation or exponentially precise tomography.

In this paper, we isolate and refine one of the simplest (but most useful) pieces of that picture: direct estimation of spectral gaps from imaginary time. Rather than reconstructing the full density of states, we ask how much can be learned about $\Delta E$ from a single imaginary-time trajectory and expectation values of local observables. We show that, for any finite-dimensional Hamiltonian $H$, it is possible to read off the spectral gap as a simple ratio of two expectation values evaluated along an imaginary-time trajectory. This can be achieved without requiring the resolution of individual eigenstates, or phase estimation. 

The resulting estimator has several desirable properties. It is agnostic to degeneracies in higher excited levels thanks to the projector formulation of the argument, and it can be extended to access $E_2 - E_1$ and higher gaps by combining logarithmic fits with additional commutator orders. Because it only involves local nested commutators with $H$, it meshes naturally with both classical simulations (where $H$ is implemented via sparse linear algebra) and quantum simulations (where $H$ is realised as a sum of few-qubit terms). It can be implemented on quantum devices using the same ITQDE machinery that underlies our earlier work \cite{leamer_quantum_2024,McCaul2025FreeSnacks}, as well as the broader QITE literature \cite{motta_determining_2020,kamakari_digital_2022,turro_imaginary-time_2022,McMahon2025QITEFlows,Alipour2025StateBasedITE}. In the context of effective-Hamiltonian approaches to differential equations \cite{Wu2025EffectiveHamiltonians,Joseph2020KvN,NovikauJoseph2025AdvecDiff,NovikauJoseph2025Carleman,Joseph2023Fusion,ParkerJoseph2020GenEigQPE}, the same construction offers a way to estimate relaxation rates and spectral gaps of auxiliary generators using only imaginary-time evolution and local measurements.

The rest of the paper is organised as follows. In Sec.~\ref{sec:method} we review imaginary time propagation and derive the commutator-based gap estimator, including its behaviour in the presence of degeneracies and its extension to higher gaps. In Secs.~\ref{sec:tfim} and \ref{sec:fh} we benchmark the method on the 1D transverse-field Ising and Fermi–Hubbard models using classical simulations. In Sec.~\ref{sec:qde} we describe how to implement the estimator using ITQDE on quantum hardware, analyse the associated measurement and quadrature costs, and compare the tradeoffs to those of full spectral reconstruction within the same framework. We conclude in Sec.~\ref{sec:conclusion} with a discussion of how this gap-focused perspective dovetails with the broader imaginary-time and effective-Hamiltonian programmes outlined above.

\section{\label{sec:method}Method}

Imaginary time propagation is a standard route to ground states and low-lying spectra \cite{chin_any_2009, lehtovaara_solution_2007}. The core idea is that as a wavefunction evolves in imaginary time $\tau$, higher-energy components are exponentially suppressed relative to lower-energy ones. In the limit $\tau \to \infty$, and provided the initial state has some overlap with the ground-state manifold, the dynamics project onto that manifold. In this section we show how a slightly more careful use of this same structure allows one to extract the \emph{gap} between the ground and first excited energies from expectation values of nested commutators of $H$ with a local observable $O$.

Let $H$ be a finite-dimensional Hamiltonian with spectral decomposition
\begin{align}
    H = \sum_{n=0}^N E_n \Pi_n, \label{eq:h_decomp}
\end{align}
where $E_0 < E_1 < \dots < E_N$ are distinct eigenvalues and the $\Pi_n$ are orthogonal projectors, $\Pi_n \Pi_m = \delta_{nm} \Pi_n$. This formulation automatically includes the case of degenerate levels: a degenerate eigenspace with eigenvalue $E_n$ is described by a projector $\Pi_n$ of rank larger than one. For an initial state $\ket{\phi_0}$, imaginary time evolution produces
\begin{align}
    \ket{\psi(\tau)} = e^{-\tau H} \ket{\phi_0}
    = \sum_{n=0}^N e^{-\tau E_n} \Pi_n \ket{\phi_0}.
\end{align}
Apart from an overall norm, the coefficients of higher-energy components decay faster in $\tau$, so that for large $\tau$ the state is dominated by contributions from the lowest few eigenspaces.

We will be interested in expectation values of operators of the form $[H,O]_M$, the $M$-th nested commutator of $H$ and a ``coordinating'' observable $O$,
\begin{align}
    [H,O]_M := [\underbrace{H, [H, \ldots, [H}_{\mbox{$M$-times}}, O]]\ldots],
\end{align}
evaluated in the imaginary-time-evolved state,
\begin{align}
    \langle [H,O]_M \rangle_\tau
    := \bra{\psi(\tau)} [H,O]_M \ket{\psi(\tau)}.
\end{align}
The key observation is that inserting the spectral decomposition of $H$ into these expressions reveals explicit powers of eigenvalue differences $(E_l - E_k)$, which can be isolated by taking ratios of expectations at different commutator orders.

We now state and prove the main result for imaginary-time evolution under $H$.

\begin{thm}\label{method}
   Let $H$ be a finite-dimensional Hamiltonian with spectral decomposition
   \begin{align}
       H = \sum_{n=0}^N E_n \Pi_n, \tag{\ref{eq:h_decomp}}
   \end{align}
   where the $\Pi_n$ are orthogonal projectors and $E_0 < E_1 < \dots < E_N$ are distinct eigenvalues. Let
   \begin{align}
       \ket{\psi(\tau)} = e^{-\tau H} \ket{\phi_0}
       = \sum_{n=0}^N e^{-\tau E_n} \Pi_n \ket{\phi_0} \label{eq:imag_time_state}
   \end{align}
   be the imaginary-time-evolved state for some initial $\ket{\phi_0}$. Define the $M$-th nested commutator
   \begin{align}
       [H,O]_M = [\underbrace{H, [H, \ldots, [H}_{\mbox{$M$-times}}, O]]\ldots],
   \end{align}
   for a coordinating observable $O$. Suppose that:
   \begin{enumerate}
       \item $\Pi_0 \ket{\phi_0} \neq 0$ and $\Pi_1 \ket{\phi_0} \neq 0$ (the initial state has support on both the ground and first excited subspaces);
       \item $\bra{\phi_0}\Pi_0 O \Pi_1 \ket{\phi_0} \neq 0$ (the observable $O$ couples these subspaces);
       \item $[H,O]_M \neq 0$ and $[H,O]_{M+2} \neq 0$.
   \end{enumerate}
   Then, as $\tau \to \infty$,
   \begin{align}
       \frac{\langle[H,O]_{M+2}\rangle_\tau}{\langle[H,O]_{M}\rangle_\tau}
       = (E_0 - E_1)^2 + \mathcal{O}\!\big(e^{-\tau(E_2 - E_1)}\big),
       \label{eq:comm_spectral_gap}
   \end{align}
   where $\langle[H,O]_{M}\rangle_\tau \equiv \bra{\psi(\tau)}[H,O]_M\ket{\psi(\tau)}$.
\end{thm}

\begin{proof}
Consider the expectation value of $[H,O]_M$,
\begin{align}
    \langle[H,O]_M\rangle_\tau
    &= \bra{\psi(\tau)}[H,O]_M\ket{\psi(\tau)} \notag \\
    &= \sum_{l,k=0}^N e^{-\tau(E_l+E_k)}
       \bra{\phi_0}\Pi_l[H,O]_M\Pi_k\ket{\phi_0}. \label{eq:comm_expt}
\end{align}
Using the expansion of the $M$-th nested commutator \cite{volkin_iterated_nodate},
\begin{align}
    [H,O]_M = \sum_{m=0}^M (-1)^m \binom{M}{m} H^{M-m} O H^m,
    \label{eq:comm_expansion}
\end{align}
(see App.~\ref{app:comm} for a proof), we obtain
\begin{align}
    \langle[H,O]_M\rangle_\tau
    &= \sum_{l,k=0}^N e^{-\tau(E_l+E_k)} \sum_{m=0}^M (-1)^m \binom{M}{m} \notag \\
    &\qquad\times \bra{\phi_0}\Pi_l H^{M-m} O H^m \Pi_k\ket{\phi_0}.
\end{align}
Applying the orthogonality of the projectors, $\Pi_n \Pi_m = \delta_{nm}\Pi_n$, yields
\begin{align}
    \langle[H,O]_M\rangle_\tau
    &= \sum_{l,k=0}^N \sum_{m=0}^M e^{-\tau(E_l+E_k)} (-1)^m \binom{M}{m} \notag \\
    &\qquad\times E_l^{\,M-m} E_k^{\,m}\, \bra{\phi_0}\Pi_l O \Pi_k\ket{\phi_0}.
\end{align}
The inner sum over $m$ is the binomial expansion of $(E_l - E_k)^M$, so we can rewrite this as
\begin{align}
    \langle[H,O]_M\rangle_\tau
    &= \sum_{l,k=0}^N e^{-\tau(E_l+E_k)}
       (E_l - E_k)^M \bra{\phi_0}\Pi_l O \Pi_k\ket{\phi_0}.
       \label{eq:sum_expt}
\end{align}
Terms with $l = k$ vanish because $(E_l - E_l)^M = 0$. For large $\tau$ the dominant contributions come from the smallest values of $E_l + E_k$, i.e. from pairs involving the ground and first excited energies. Retaining only the $l,k \in \{0,1\}$ terms gives
\begin{align}
    \langle[H,O]_M\rangle_\tau
    &= e^{-\tau(E_0+E_1)} (E_0 - E_1)^M \times \notag \\
    &\qquad\Big[
      \bra{\phi_0}\Pi_0 O \Pi_1\ket{\phi_0}
      + (-1)^M \bra{\phi_0}\Pi_1 O \Pi_0\ket{\phi_0}
    \Big] \notag \\
    &\quad + \mathcal{O}\!\big(e^{-\tau(E_0+E_2)}\big),
    \label{eq:mth_comm}
\end{align}
where $\mathcal{O}(e^{-\tau(E_0+E_2)})$ collects all contributions involving the second excited energy $E_2$ and above. Note that the bracket in Eq.~\eqref{eq:mth_comm} can be written as $2 \Re A$ when $M$ is even, and $2 i \Im A$ when $M$ is odd, with
$A=\bra{\phi_0}\Pi_0 O \Pi_1\ket{\phi_0}$. Our assumption that $O$
couples the ground and first excited subspaces is precisely the
requirement that this prefactor does not vanish. The same reasoning applied to $M+2$ yields
\begin{align}
    \langle[H,O]_{M+2}\rangle_\tau
    &= e^{-\tau(E_0+E_1)} (E_0 - E_1)^{M+2} \times \notag \\
    &\qquad\Big[
      \bra{\phi_0}\Pi_0 O \Pi_1\ket{\phi_0}
      + (-1)^{M+2} \bra{\phi_0}\Pi_1 O \Pi_0\ket{\phi_0}
    \Big] \notag \\
    &\quad + \mathcal{O}\!\big(e^{-\tau(E_0+E_2)}\big).
\end{align}
Taking the ratio of $\langle[H,O]_{M+2}\rangle_\tau$ and $\langle[H,O]_M\rangle_\tau$, the common exponential factor $e^{-\tau(E_0+E_1)}$ and the common prefactor in square brackets cancel, leaving
\begin{align}
    \frac{\langle[H,O]_{M+2}\rangle_\tau}{\langle[H,O]_{M}\rangle_\tau}
    &= (E_0 - E_1)^2
       + \mathcal{O}\!\big(e^{-\tau(E_2 - E_1)}\big),
\end{align}
where we have expressed the leading correction in terms of the next gap $E_2 - E_1$. This proves Eq.~\eqref{eq:comm_spectral_gap}.
\end{proof}

The use of projectors in Eq.~\eqref{eq:h_decomp} makes the treatment of degeneracies straightforward. For a three-level system with non-degenerate energies one has $H = E_0 \Pi_0 + E_1 \Pi_1 + E_2 \Pi_2$ with $\Pi_n = \ket{n}\bra{n}$. If instead the ground state is two-fold degenerate, one can take $\Pi_0 = \ket{0}\bra{0} + \ket{1}\bra{1}$ and $\Pi_1 = \ket{2}\bra{2}$, with $H = E_0 \Pi_0 + E_1 \Pi_1$. In both cases Eq.~\eqref{eq:comm_spectral_gap} yields the same gap $\Delta E = E_1 - E_0$, provided $O$ connects the relevant subspaces.

\subsection*{Logarithmic estimators and higher gaps}

The ratio in Eq.~\eqref{eq:comm_spectral_gap} is not the only way to access spectral information. Taking the modulus and logarithm of Eq.~\eqref{eq:mth_comm} yields
\begin{align}
    \ln\big|\langle[H,O]_{M}\rangle_\tau\big|
    = -\tau(E_0 + E_1) + \mathcal{O}(1),
    \qquad \tau \to \infty.
    \label{eq:ln_comm}
\end{align}
Thus, a linear fit of $\ln|\langle[H,O]_M\rangle_\tau|$ versus $\tau$ at large $\tau$ yields $-(E_0 + E_1)$ as its slope, up to corrections that do not grow with $\tau$. If $E_0$ is known independently (for example from a separate imaginary-time ground-state calculation), this provides an alternative way to obtain $\Delta E = E_1 - E_0$ without forming the ratio in Eq.~\eqref{eq:comm_spectral_gap}. This can be advantageous in situations where $\langle[H,O]_{M+2}\rangle_\tau$ and $\langle[H,O]_{M}\rangle_\tau$ are both small and their ratio is numerically delicate.

The same logic can be used to access the second gap $E_2 - E_1$. Rearranging Eq.~\eqref{eq:comm_spectral_gap},
\begin{align}
    \left|\frac{\langle[H,O]_{M+2}\rangle_\tau}{\langle[H,O]_{M}\rangle_\tau}
          - \Delta E^2 \right|
    = \mathcal{O}\!\big(e^{-\tau(E_2 - E_1)}\big),
\end{align}
so that
\begin{align}
    \ln\left| \frac{\langle[H,O]_{M+2}\rangle_\tau}{\langle[H,O]_{M}\rangle_\tau}
               - \Delta E^2 \right|
    = -\tau(E_2 - E_1) + \mathcal{O}(1),
    \label{eq:e2e1gap}
\end{align}
where $\Delta E = E_1 - E_0$ is obtained from Eq.~\eqref{eq:comm_spectral_gap} using data at the largest available $\tau$. Thus, a second linear fit of the residual on a logarithmic scale yields the next gap $E_2 - E_1$. In principle one can continue this process, using higher-order commutators (e.g. $M+4$ and above) to construct systems of polynomial equations whose coefficients depend on combinations of eigenvalues, and solve these to ladder up the spectrum. In practice, we focus on the lowest and second gaps, where the numerical behaviour is most favourable.

The constructions above are formulated for direct imaginary-time evolution under $H$. In Sec.~\ref{sec:qde} we show how the same estimators can be implemented when imaginary time is accessed via ITQDE—i.e., when $e^{-\tau H^2}$ is realised as a finite linear combination of short-time unitaries—and derive the corresponding version of Theorem~\ref{method} for that setting.

\section{\label{sec:results}Results}

We now illustrate the commutator-based estimators developed in Sec.~\ref{sec:method} on two standard model systems: the one-dimensional transverse-field Ising model and the one-dimensional Fermi–Hubbard model. In each case we compare the gap $\Delta E$ obtained from Eq.~\eqref{eq:comm_spectral_gap} against the exact result from full diagonalisation, and we examine the logarithmic estimators for $E_0 + E_1$ and the second gap $E_2 - E_1$ derived in Eqs.~\eqref{eq:ln_comm} and \eqref{eq:e2e1gap}.

Throughout, we quantify the accuracy of a given estimate $\Delta E_{\mathrm{method}}$ via the relative error
\begin{align}
    \epsilon
    = \frac{|\Delta E_{\mathrm{exact}} - \Delta E_{\mathrm{method}}|}
           {|\Delta E_{\mathrm{exact}}|},
    \label{eq:rel_err}
\end{align}
where $\Delta E_{\mathrm{exact}}$ is obtained from exact diagonalisation of the relevant finite-size Hamiltonian. All classical simulations are carried out using the QuSpin library \cite{quspin_git}. The code used to generate the data presented here is available in Ref.~\cite{leamer_git}.

\subsection{\label{sec:tfim}Transverse-field Ising model}

We first consider the one-dimensional transverse-field Ising model (TFIM), a paradigmatic testbed for quantum phase transitions \cite{dziarmaga_dynamics_2005,sachdev_quantum_1999}, spin glasses \cite{kopec_instabilities_1989,laumann_cavity_2008}, and quantum annealing dynamics \cite{farhi_quantum_2000,troels_2014,shin_how_2014,boixo_evidence_2014}. The Hamiltonian reads
\begin{align}
    H = -J\sum_{l=0}^{L-1}Z_{l}Z_{l+1} - h\sum_{l=0}^{L-1}X_l,
    \label{eq:tfim}
\end{align}
where $Z_l$ and $X_l$ are Pauli operators on site $l$, $J$ is the nearest-neighbour interaction strength, $h$ is the transverse field, and $L$ is the number of lattice sites. We impose periodic boundary conditions, so that $Z_L \equiv Z_0$. For the system sizes considered here the model is easily diagonalised, which allows direct comparison with our estimator.

A central practical ingredient in the method is the choice of coordinating observable $O$. For the TFIM we take
\begin{equation}
    O = -J (Z_0 + Z_1),
\end{equation}
which corresponds to the (scaled) local magnetisation on two adjacent sites. This choice is local and corresponds to the magnetisation on two adjacent sites. It is also odd under the global $\mathbb{Z}_2$ parity symmetry $P = \prod_l X_l$, so it connects states in opposite parity sectors; for the parameters considered, the ground and first excited states lie in different sectors, ensuring that $O$ has non-vanishing matrix elements between them. The overall factor of $J$ simply makes the dimensions consistent with the Hamiltonian; it cancels in the ratio Eq.~\eqref{eq:comm_spectral_gap} and therefore does not affect the estimated gap.

As an initial state $\ket{\phi_0}$ we use a normalised random complex vector in the computational basis, with real and imaginary parts of each component drawn independently from a uniform distribution on $[0,1]$. This construction has high probability of yielding non-zero overlap with both the ground and first excited eigenspaces; in practice we verify this by checking that the exact projections $\Pi_0 \ket{\phi_0}$ and $\Pi_1 \ket{\phi_0}$ are non-zero for the instances studied.

\begin{figure}
    \centering
    \includegraphics[width=\linewidth]{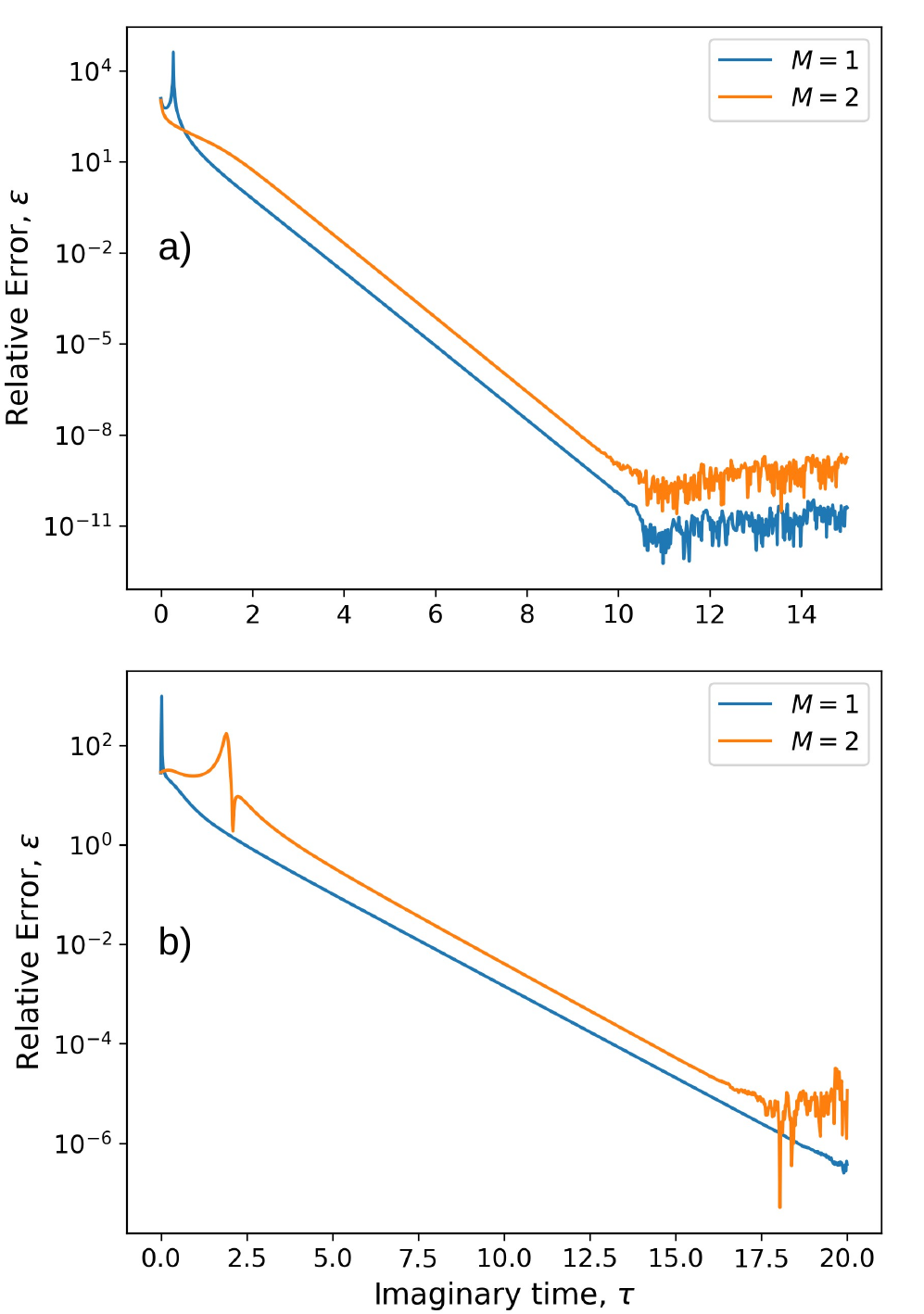}
    \caption{Relative error $\epsilon$ in $\Delta E$ computed via Eq.~\eqref{eq:comm_spectral_gap} plotted versus imaginary time $\tau$ for (a) the 1D transverse-field Ising model and (b) the 1D Fermi–Hubbard model, with $M=1$ (blue) and $M=2$ (orange). In both cases we observe an extended regime of exponential decay in $\epsilon$ before the error eventually increases again as the contribution from the first excited state is quenched.}
    \label{fig:tfim_rel_err}
\end{figure}

Figure~\ref{fig:tfim_rel_err}a) shows the relative error $\epsilon$ in the estimated spectral gap $\Delta E$ as a function of imaginary time $\tau$ for a TFIM with $L=4$, $J=1$, and $h=1$. We plot results for the first two non-trivial commutator orders, $M=1$ (blue) and $M=2$ (orange), using Eq.~\eqref{eq:comm_spectral_gap}. In both cases the relative error decreases exponentially with $\tau$ over an extended window, reaching values of order $10^{-8}$ or lower for moderate $\tau$. The negative linear trend on the semi-logarithmic scale is precisely what is expected from the error term in Eq.~\eqref{eq:comm_spectral_gap}, which predicts corrections of order $e^{-\tau(E_2 - E_1)}$.

At very large $\tau$ the error eventually increases again. From the point of view developed in Ref.~\cite{McCaul2025FreeSnacks}, this is the same finite-bandwidth tradeoff that governs the ITQDE spectral staircase \cite{leamer_quantum_2024,McCaul2025FreeSnacks}. Imaginary time defines a Gaussian spectral filter (see Sec. \ref{sec:qde}) of width $\sim 1/\sqrt{\tau}$: for small $\tau$ the filter is too broad and contributions from $E_2$ and higher levels are not yet suppressed, while for very large $\tau$ the filter is so narrow that the weight of the first excited component of $\ket{\psi(\tau)}$ becomes exponentially small compared to the ground state. In the staircase language, pushing $\tau$ too far corresponds to working in a region where the filter is so narrow that its normalisation is on the same scale as the numerical or sampling error~\cite{McCaul2025FreeSnacks}. In such windows the estimator becomes locally unstable: the observable is effectively supported only on the ground-state subspace and loses sensitivity to the gap. The exponential decay and subsequent turn-around of the relative error in Fig.~\ref{fig:tfim_rel_err}a) therefore reflect two sides of the same mechanism: imaginary time must be long enough to suppress higher excitations, but not so long that the contribution from the first excited state is no longer distinguishable from numerical or sampling noise.

We can also use the TFIM to test the logarithmic estimators for $E_0 + E_1$ and $E_2 - E_1$. Table~\ref{tab:alt_gaps}a) summarises the results for the same parameter set, using $M=1$. The quantity labelled $\Delta E_{\mathrm{method}}$ is obtained by fitting the slopes of $\ln|\langle[H,O]_1\rangle_\tau|$ and of
\begin{align}
    \ln\left| \frac{\langle[H,O]_{3}\rangle_\tau}
                     {\langle[H,O]_{1}\rangle_\tau}
               - \Delta E^2 \right|
\end{align}
versus $\tau$ at large $\tau$, as prescribed by Eqs.~\eqref{eq:ln_comm} and \eqref{eq:e2e1gap}, respectively. The agreement for $E_0 + E_1$ is at the level of $10^{-6}$ relative error, while the estimate for the second gap $E_2 - E_1$ is accurate at the few-percent level. This confirms that the logarithmic variants are viable alternatives to the ratio estimator when one wishes to avoid dividing small numbers.

\begin{table}[h]
    \centering
    \begin{tabular}{c | c | c | c}
         & $\Delta E_{\mathrm{exact}}$ & $\Delta E_{\mathrm{method}}$ & Relative error \\
         \hline
         $E_0+E_1$ & -10.05 & -10.05 & $7.4 \times 10^{-7}$\\
         $E_2-E_1$ & 2.664 & 2.804 & $5.8 \times 10^{-2}$ \\
         \hline
         \multicolumn{4}{c}{}\\
         \multicolumn{4}{c}{(a) 1D transverse-field Ising model} \\
         \multicolumn{4}{c}{}\\
         & $\Delta E_{\mathrm{exact}}$ & $\Delta E_{\mathrm{method}}$ & Relative error\\
        \hline
        $E_0+E_1$ & -5.633 & -5.632 &  $1.7 \times 10^{-4}$\\
        $E_2-E_1$ & -0.8363 & -0.8504 & $1.7 \times 10^{-2}$\\
        \hline
        \multicolumn{4}{c}{}\\
         \multicolumn{4}{c}{(b) 1D Fermi–Hubbard model} \\
    \end{tabular}
    \caption{Relative error in using Eq.~\eqref{eq:ln_comm} and Eq.~\eqref{eq:e2e1gap} to calculate $E_0+E_1$ and $E_2-E_1$, respectively, with $M=1$. In both models, the estimator for $E_0 + E_1$ is highly accurate, while the estimator for the second spectral gap is accurate at the few-percent level.}
    \label{tab:alt_gaps}
\end{table}

\subsection{\label{sec:fh}Fermi-Hubbard Model}

We next consider applications to the 1D Fermi--Hubbard (FH) model, a paradigmatic setting for strongly correlated lattice fermions. The model captures competition between kinetic delocalisation and on-site repulsion, and has been used as a minimal framework for Mott transitions, unconventional superconductivity, and magnetism in both condensed–matter and ultracold–atom contexts \cite{imada_1998,anderson_1987,hofrichter_2016}. In applications where Hubbard-like models appear as effective Hamiltonians for optical or transport problems \cite{McCaul2020DrivenImposters,McCaul2020ControllingArbitrary}, the ability to estimate $\Delta E$ and $E_2-E_1$ in this way offers a route to quantifying relaxation rates and separation of timescales that underlie observed behaviours. In Sec.~\ref{sec:qde} we show that the same construction can be ported to quantum hardware using ITQDE, retaining this connection between imaginary-time relaxation and effective spectral structure.

The FH Hamiltonian reads
\begin{align}
    H &= -t \sum_{l=0, \sigma}^{L-1}\big(c_{l\sigma}^\dag c_{l+1,\sigma} - c_{l\sigma}c_{l+1,\sigma}^\dag\big)
    +U\sum_{l=0}^{L-1}n_{l\uparrow}n_{l\downarrow},
    \label{eq:fh_model}
\end{align}
where $c_{l\sigma}^\dag, c_{l\sigma}$ are fermionic creation and annihilation operators at site $l$ with spin $\sigma$, $n_{l\sigma}=c_{l\sigma}^\dag c_{l\sigma}$, $t$ denotes the tunnelling amplitude, and $U$ the on-site interaction strength. We again impose periodic boundary conditions. As a coordinating observable we take
\begin{equation}
    O = n_{0} + n_{1},
\end{equation}
where $n_l = n_{l\uparrow} + n_{l\downarrow}$ is the local occupation on site $l$. This is a simple local density probe, and for the chain lengths and parameters considered it satisfies the commutator conditions of Theorem~\ref{method}: $[H,O]_M$ and $[H,O]_{M+2}$ are non-vanishing and the matrix element $\bra{\phi_0}\Pi_0 O \Pi_1 \ket{\phi_0}$ is generically non-zero and complex for a random initial state.

As in Sec.~\ref{sec:tfim}, we benchmark the gap estimator by computing the relative error of Eq.\eqref{eq:rel_err}, where $\Delta E_{\mathrm{exact}}$ is obtained by direct diagonalisation of Eq.~\eqref{eq:fh_model} and $\Delta E_{\mathrm{method}}$ from Eq.~\eqref{eq:comm_spectral_gap}. We consider a half-filled chain with $L=4$, $t=1$, and $U=\sqrt{2}$, propagated in imaginary time using QuSpin \cite{quspin_git}. The initial state $\ket{\phi_0}$ is again chosen as a random complex vector with independently and uniformly distributed real and imaginary components, normalised to unity.

Figure~\ref{fig:tfim_rel_err}b) shows $\epsilon(\tau)$ for $M=1$ and $M=2$. As in the transverse-field Ising case, the error initially decays approximately exponentially in $\tau$, in line with the $e^{-\tau(E_2-E_1)}$ correction in Eq.~\eqref{eq:comm_spectral_gap}, before turning around once the first excited component has been overly suppressed. For both commutator orders the relative error reaches $\sim 10^{-4}$ or better over an extended window of imaginary times, indicating that the method remains accurate in an interacting fermionic setting where the low-energy manifold carries nontrivial spin and charge structure.

We also test the alternative estimators based on logarithmic fits. Using the same parameters and $M=1$, we extract $E_0+E_1$ from the slope of Eq.~\eqref{eq:ln_comm} and $E_2-E_1$ from Eq.~\eqref{eq:e2e1gap}, with results summarised in Table~\ref{tab:alt_gaps}b). The error in $E_0+E_1$ remains at the $10^{-4}$ level, while $E_2-E_1$ is captured within a few percent. As in the spin-chain case, this suggests that once a reasonable estimate of $\Delta E$ has been obtained from the commutator ratio, the same imaginary-time data can be repurposed to ladder up the spectrum by combining additional commutator orders with logarithmic fits.

\section{Towards Quantum Computer Implementations}\label{sec:qde}

So far we have treated imaginary-time propagation as a classical primitive. On a quantum device, dynamics are unitary, and $e^{-\tau H}$ is not directly available. In this section we show how the same commutator-based gap estimator can be implemented on gate-based quantum hardware using Imaginary-Time Quantum Dynamical Emulation (ITQDE)~\cite{leamer_quantum_2024}. In the interests of simplicity, we forego the methodological refinements used in Ref.~\cite{McCaul2025FreeSnacks}, and describe a deliberately minimal implementation. This should therefore be read as a proof of principle, rather than an optimised hardware proposal.

Our starting point is a form of \textit{Hubbard--Stratonovich} transformation~\cite{AltlandSimons,NegeleOrland},
\begin{align}
    e^{-\tau H^2}
    = \frac{1}{\sqrt{4\pi \tau}}
      \int_{-\infty}^{\infty}\! \mathrm{d}t\,
      e^{-t^2/(4\tau)}\, e^{-i t H},
    \label{eq:HS_identity}
\end{align}
which expresses the $H^2$ kernel as a Gaussian average of real-time unitaries $e^{-i t H}$. Closely related constructions appear in functional-integral and influence-functional treatments of interacting and open quantum systems, where quadratic couplings are linearised at the cost of introducing auxiliary stochastic fields~\cite{mccaul_partition-free_2017,mccaul_how_2021}. Changing variables to $t = 2\sqrt{\tau}\, x$ yields
\begin{align}
    e^{-\tau H^2}
    = \frac{1}{\sqrt{\pi}}
      \int_{-\infty}^{\infty}\! \mathrm{d}x\,
      e^{-x^2}\, e^{-2 i \sqrt{\tau}\, x H},
    \label{eq:HS_gaussian_x}
\end{align}
which is the kernel derived from ITQDE~\cite{leamer_quantum_2024} and employed in the subsequent ``free snacks'' analysis~\cite{McCaul2025FreeSnacks}. For Hermitian $H$, the Gaussian weight is real and even in $x$, so only the even (cosine) part of the unitary contributes:
\begin{align}
    e^{-\tau H^2}
    &= \frac{1}{\sqrt{\pi}}
      \int_{0}^{\infty}\! \mathrm{d}x\,
      e^{-x^2} \cos\!\big(2 \sqrt{\tau}\, x H\big) \notag \\
    &= \frac{2}{\sqrt{\pi}}
      \int_{0}^{\infty}\! \mathrm{d}x\,
      e^{-x^2} \,\mathrm{Re}\!\big(e^{-2 i \sqrt{\tau}\, x H}\big).
    \label{eq:HS_cos}
\end{align}

Approximating the Gaussian integral with Gauss–Hermite quadrature,
\begin{align}
    \int_{-\infty}^{\infty}\! \mathrm{d}x\,
    e^{-x^2} f(x)
    \approx \sum_{k=1}^{K} w_k f(x_k),
\end{align}
gives the discrete operator approximation
\begin{align}
    e^{-\tau H^2}
    &\approx \frac{1}{\sqrt{\pi}} \sum_{k=1}^{K} w_k
        \cos\!\big(2 \sqrt{\tau}\, x_k H\big) \notag \\
    &= \frac{1}{\sqrt{\pi}} \sum_{k=1}^{K} w_k \,
        \mathrm{Re}\!\big(U_k\big),
    \qquad
    U_k \equiv e^{-2 i \sqrt{\tau}\, x_k H}.
    \label{eq:imag_time_h_sq}
\end{align}
This is the critical point of departure: the Euclidean kernel is approximated by a weighted sum of \emph{Hermitian} operators $\cos(2\sqrt{\tau}x_k H)$, or equivalently by the real parts of a finite set of unitaries $U_k$. We never approximate $e^{-\tau H^2}$ by a bare sum of unitaries without taking the real part.

Within the developed methodology, replacing $e^{-\tau H}$ by $e^{-\tau H^2}$ changes only how higher levels are suppressed; the structure of the commutator-based estimator from Sec.~\ref{sec:method} is otherwise unchanged. For completeness we state the analogue of Theorem~\ref{method}.
Replacing $e^{-\tau H}$ by $e^{-\tau H^2}$ changes only the imaginary-time weights in the spectral decomposition. As a direct corollary of Theorem~\ref{method} we have:

\begin{thm}[Imaginary time under $H^2$]\label{QDE_method}
   Let $H$, $\{\Pi_n\}$, $\ket{\phi_0}$, and $O$ satisfy the assumptions of Theorem~\ref{method} and $E_0 \geq 0$. Define
   \begin{align}
       \ket{\psi(\tau)}
       = e^{-\tau H^2} \ket{\phi_0}
       = \sum_{n=0}^N e^{-\tau E_n^2} \Pi_n \ket{\phi_0},
   \end{align}
   and $\langle[H,O]_M\rangle_\tau \equiv \bra{\psi(\tau)}[H,O]_M\ket{\psi(\tau)}$. Then, as $\tau \to \infty$,
   \begin{align}
       \frac{\langle[H,O]_{M+2}\rangle_\tau}{\langle[H,O]_M\rangle_\tau}
       = (E_1 - E_0)^2
         + \mathcal{O}\!\big(e^{-\tau (E_2^2 - E_1^2)}\big).
       \label{eq:qde_spectral_gap}
   \end{align}
\end{thm}
The proof is identical to that of Theorem~\ref{method}, with $E_n$ replaced by $E_n^2$ in the imaginary-time weights, and is therefore omitted.

Formally, the gap estimator uses
\begin{align}
    \langle[H,O]_M\rangle_\tau
    = \bra{\phi_0} e^{-\tau H^2} [H,O]_M e^{-\tau H^2} \ket{\phi_0},
    \label{eq:comm_expt_QDE_formal}
\end{align}
when Eq.~\eqref{eq:imag_time_h_sq} is substituted on both sides, it gives
\begin{align}
    \langle[H,O]_M\rangle_\tau
    &\approx \frac{1}{\pi}
       \sum_{j,k=1}^{K} w_j w_k\,
       \bra{\phi_0}
          \mathrm{Re}(U_j)\,[H,O]_M\,\mathrm{Re}(U_k)
       \ket{\phi_0}.
    \label{eq:comm_expt_QDE_real}
\end{align}
Writing
\begin{align}
    \mathrm{Re}(U_k) = \frac{1}{2}\big(U_k + U_k^\dagger\big),
    \qquad
    U_k = e^{-2 i \sqrt{\tau}\, x_k H},
\end{align}
and expanding Eq.~\eqref{eq:comm_expt_QDE_real} yields a linear combination of terms of the form
\begin{align}
    \bra{\phi_0} U_j^{s} [H,O]_M U_k^{s'} \ket{\phi_0},
    \qquad s,s'\in\{+1,-1\},
\end{align}
with real coefficients determined entirely by the Gauss–Hermite weights $\{w_k\}$. We can choose the nodes and weights in symmetric pairs $\{\pm x_k, w_k\}$, so all these coefficients are real. To make contact with quantities that are natural to measure, we expand the commutator in the Pauli basis,
\begin{align}
    [H,O]_M = \sum_{\alpha=1}^{R} c_\alpha P_\alpha,
    \label{eq:pauli_expansion}
\end{align}
with $c_\alpha \in \mathbb{C}$ and each $P_\alpha$ a tensor product of Pauli matrices. Inserting Eq.~\eqref{eq:pauli_expansion} into Eq.~\eqref{eq:comm_expt_QDE_real} and regrouping terms gives
\begin{align}
    \langle[H,O]_M\rangle_\tau
    \approx \frac{1}{\pi}
    \sum_{j,k=1}^{K} \sum_{\alpha=1}^{R}
    w_j w_k\, c_\alpha\,
    \mathrm{Re}\!\left(
        \bra{\phi_0} U_j^\dagger P_\alpha U_k \ket{\phi_0}
    \right),
    \label{eq:comm_expt_Pauli}
\end{align}
where the appearance of the real part is now explicit: it arises from the even HS kernel and the symmetric quadrature, not from any ad hoc choice.

Moreover, because the kernel is real and even, and the weights $w_j w_k$ are real, we may, without loss of generality, choose the $c_\alpha$ to be real by absorbing phases into the $P_\alpha$ strings. In that case the only complex phases in each term come from the overlaps
\begin{align}
    \bra{\phi_0} U_j^\dagger P_\alpha U_k \ket{\phi_0}.
\end{align}
Combining the contributions from $(j,k)$ and $(k,j)$ and using
$U_j^\dagger P_\alpha U_k = (U_k^\dagger P_\alpha U_j)^\dagger$ shows that the imaginary parts cancel pairwise in the double sum. Equivalently, one may restrict to $j \le k$ and write
\begin{align}
    \langle[H,O]_M\rangle_\tau
    \approx \frac{2}{\pi}
    \sum_{j \le k} \sum_{\alpha=1}^{R}
    w_j w_k\, c_\alpha\,
    \mathrm{Re}\!\left(
        \bra{\phi_0} U_j^\dagger P_\alpha U_k \ket{\phi_0}
    \right).
    \label{eq:comm_expt_Pauli_real}
\end{align}
Thus the HS + Gauss–Hermite structure has already projected the estimator onto the real parts of these overlaps. This means that it is only  necessary to obtain $\mathrm{Re}\!\left(\bra{\phi_0} U_j^\dagger P_\alpha U_k \ket{\phi_0}\right)$ from circuit calculation.

For Hermitian $H$ and $O$, the operator $[H,O]_M$ is anti-Hermitian and its expectation values are purely imaginary. In practice we implement Eq.~\eqref{eq:pauli_expansion} with purely imaginary coefficients $c_\alpha = i \tilde{c}_\alpha$, $\tilde{c}_\alpha \in \mathbb{R}$, so that the product of $i$ with the real overlaps in Eq.~\eqref{eq:comm_expt_Pauli_real} yields a purely imaginary $\langle[H,O]_M\rangle_\tau$ as required.

Each overlap
\begin{align}
    \mathrm{Re}\!\left(
        \bra{\phi_0} U_j^\dagger P_\alpha U_k \ket{\phi_0}
    \right)
\end{align}
can then be estimated using standard circuit primitives. Thanks to Eq.~\eqref{eq:comm_expt_Pauli_real}, only its real part ever contributes: the HS kernel and symmetric quadrature guarantee that imaginary components cancel in the final estimator.

This matches exactly what the simplest version of the Hadamard test~\cite{cleve_quantum_1998} returns. An ancilla qubit is initialised in $\ket{0}$ and the system register in $\ket{\phi_0}$. A Hadamard gate is applied to the ancilla, followed by a controlled unitary that implements $U_j^\dagger P_\alpha U_k$ on the system, conditioned on the ancilla being in $\ket{1}$. A second Hadamard is then applied to the ancilla, followed by a measurement in the computational basis. The probability $p_0$ of outcome $0$ on the ancilla is related to the real part of the overlap by
\begin{align}
    \mathrm{Re}\!\left(\bra{\phi_0} U_j^\dagger P_\alpha U_k \ket{\phi_0}\right)
    = 2 p_0 - 1.
\end{align}
If one wished to access the imaginary part, a phase gate on the ancilla between the first Hadamard and the controlled unitary would suffice, but the HS symmetry means this is never needed for the gap estimator.

To illustrate how the gap estimator behaves under this minimal ITQDE implementation, we simulate a simple example using Qiskit’s Aer backend~\cite{Qiskit}. We consider the 1D transverse-field Ising model with $L=2$, $J=1$, and $h=1$, as in Sec.~\ref{sec:tfim}. The initial state $\ket{\phi_0}$ is taken to be a normalised random complex vector in the two-qubit computational basis, with independently and uniformly distributed real and imaginary parts, matching the classical tests.

As a coordinating observable we choose
\begin{align}
    O = Y \otimes I + I \otimes Y,
\end{align}
where $Y$ is the Pauli-$Y$ operator and $I$ is the identity. This operator is local, changes the global $Z$-parity, and for the chosen parameters has non-vanishing matrix elements between the ground and first excited states. The commutators $[H,O]_M$ are constructed explicitly and decomposed into Pauli strings. We use Eq.~\eqref{eq:imag_time_h_sq} with a fixed Gauss–Hermite order $K=30$ for all values of $\tau$. The unitaries $U_k = e^{-2 i \sqrt{\tau}\, x_k H}$ are implemented using Qiskit’s built-in Hamiltonian simulation tools, without attempting to optimise Trotter steps or gate decompositions. To isolate sampling effects, we simulate an otherwise ideal device and include only shot noise, using $10^6$ shots for each Hadamard-test configuration.

\begin{figure}
    \centering
    \includegraphics[width=\linewidth]{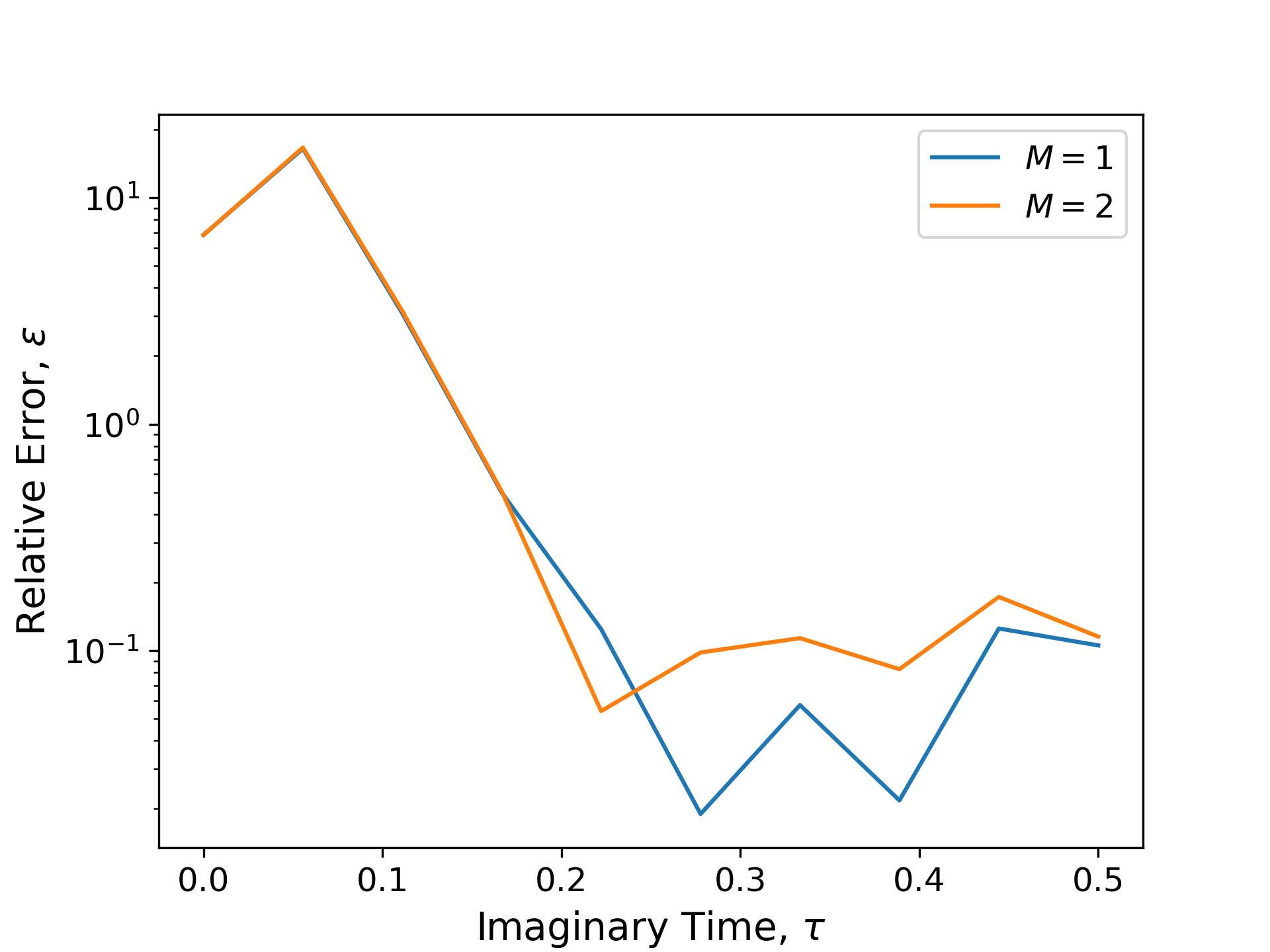}
    \caption{Relative error $\epsilon$ in $\Delta E$ computed via Eq.~\eqref{eq:qde_spectral_gap} plotted versus imaginary time $\tau$ for the 1D transverse-field Ising model with $L=2$, using $M=1$ (blue) and $M=2$ (orange). Calculations are performed using Qiskit’s Aer simulator with $10^6$ shots per Hadamard-test configuration and $K=30$ Gauss–Hermite points.}
    \label{fig:qde_aer_sim}
\end{figure}

The resulting relative error $\epsilon$ in the extracted gap as a function of $\tau$ is shown in Fig.~\ref{fig:qde_aer_sim} for $M=1$ and $M=2$, using Eq.~\eqref{eq:qde_spectral_gap}. As in the exact simulations, there is an intermediate ``Goldilocks'' window in $\tau$ where higher excitations are suppressed but the first excited contribution remains visible. However, the minimum error reached in these simulations is only at the level of $\epsilon \sim 10^{-1}$, substantially worse than the classical benchmarks in Fig.~\ref{fig:tfim_rel_err}.

The behaviour observed under this crude implementation is entirely consistent with the bandwidth analysis of ITQDE developed in Ref.~\cite{McCaul2025FreeSnacks}. The same Gaussian kernel that cleanly suppresses higher excited states also drives the relevant overlaps $\bra{\phi_0} U_j^\dagger P_\alpha U_k \ket{\phi_0}$ towards the sampling floor if $\tau$ and $K$ are pushed without regard for the available shot budget. Once these overlaps fall to the scale of the shot noise, the estimates of $\langle[H,O]_M\rangle_\tau$ and $\langle[H,O]_{M+2}\rangle_\tau$ are dominated by statistical fluctuations, and forming their ratio amplifies the effect.

This section should therefore be read as a proof-of-principle: the commutator-based gap estimator is fully compatible with the ITQDE / ``free snacks'' machinery, and can be expressed entirely in terms of the same HS kernel, Gauss–Hermite discretisation, and overlap primitives. Making it competitive in practice would require importing the full bandwidth analysis of Ref.~\cite{McCaul2025FreeSnacks} to choose $(\tau,K)$, designing coordinating observables $O$ that maximise the relevant matrix elements, and replacing the brute-force Hadamard tests by more measurement-efficient overlap estimators. Such extensions are straightforward in principle, but reserved for future work. 

\section{\label{sec:conclusion}Conclusion and outlook}
In this work we have developed a construction for estimating the spectral gap, $\Delta E = E_1 - E_0$, associated with a finite-dimensional Hamiltonian from imaginary-time dynamics. This is done by monitoring a single family of nested commutators with a suitably chosen local ``coordinating'' observable. The central object is the ratio $\frac{\langle[H,O]_{M+2}\rangle_\tau}{\langle[H,O]_M\rangle_\tau}$, evaluated along an imaginary-time trajectory. Under mild assumptions on the initial state and the coordinating observable $O$, this ratio converges to $(E_0 - E_1)^2$ with corrections controlled by the next excited manifold. Importantly, the argument is formulated in terms of projectors $\Pi_n$ rather than individual eigenvectors, so it applies equally in the presence of degeneracies in higher levels.

Classical benchmarks on the 1D transverse-field Ising and Fermi–Hubbard models validate the construction. In both cases, the gap estimate converges rapidly as a function of imaginary time, with an error profile controlled by the next gap in the spectrum. The breakdown at very large $\tau$—where the first excited component is itself suppressed below numerical noise—is the same mechanism that powers the method: there is a  window in which higher levels are quenched but the first excited state remains visible. Within that window the estimator achieves high accuracy without resolving individual eigenstates or invoking phase estimation. The method is deliberately modest in scope. It does not bypass the worst-case complexity of the local Hamiltonian problem, and it is not a replacement for phase estimation or for the full ITQDE machinery when a detailed density of states is required. What it does offer is a structurally simple way of turning imaginary-time propagation into a diagnostic for low-lying gaps, with costs that are easy to account for: once $H$ and $O$ are fixed, one pays for nested commutators and for propagating a single state in imaginary time. 

On quantum hardware, we show how the same estimator can be implemented within the ITQDE framework. A Hubbard–Stratonovich transformation expresses $e^{-\tau H^2}$ as a Gaussian average over real-time unitaries, discretised via Gauss–Hermite quadrature. The resulting Euclidean expectation values reduce to weighted sums of real parts of overlaps of the form $\bra{\phi_0}U_j^\dagger P_\alpha U_k\ket{\phi_0}$, which can be obtained by standard Hadamard tests. The Aer simulations reported here constitute a proof of principle rather than an optimised proposal: they make clear that bandwidth and shot-noise constraints, analysed in detail in Ref.~\cite{McCaul2025FreeSnacks}, will limit performance unless $\tau$, the quadrature order, and the choice of coordinating observable are tuned with some care. A full analysis of this, in tandem with the results obtained in Ref.~\cite{McCaul2025FreeSnacks}, constitute a central future objective. Nevertheless, we may already infer that the commutator-based estimator has a modest but genuine structural advantage. In Refs.~\cite{McCaul2025FreeSnacks,leamer_quantum_2024}, spectral information is encoded via an additional free parameter $\lambda$, which necessitates extracting the full phase information of overlaps, as well as explicit normalisation of the emulated state. Here the structure the ratio of commutator structure eliminates these additional requirements. 

The same structure also dovetails with the broader programme in which effective Hamiltonians are used as workhorses for solving differential equations. In block-encoding / LCU approaches and in Koopman–von Neumann or Liouville embeddings, auxiliary generators are engineered so that their spectra encode transport, relaxation, or fusion-relevant dynamics. For such operators the quantity of interest is often precisely a gap or low-lying relaxation rate, rather than the full spectrum. The present construction shows that, once one can implement imaginary-time evolution (classically or via ITQDE), these gaps can be accessed using only local commutators and overlap measurements.

There are several future directions for refinement. On the classical side, one can combine higher commutator orders with improved fitting strategies to stabilise estimates of multiple gaps. On the quantum side, importing the full bandwidth analysis of Ref.~\cite{McCaul2025FreeSnacks} to choose $(\tau,K)$, designing coordinating observables that maximise the relevant matrix elements, and replacing brute-force Hadamard tests by more measurement-efficient overlap estimators are natural next steps. In this case, the gap to future development is manifestly small.

\begin{acknowledgments}

We gratefully acknowledge discussions with Andrew Baczewski. J.M.L. and A.B.M. acknowledge support from Sandia National Laboratories’ Laboratory Directed Research and Development Program under the Truman Fellowship and Project 222396. Sandia National Laboratories is a multimission laboratory managed and operated by National Technology \& Engineering Solutions of Sandia, LLC, a wholly owned subsidiary of Honeywell International Inc., for the U.S. Department of Energy’s National Nuclear Security Administration under contract DE-NA0003525. This paper describes objective technical results and analysis. Any subjective views or opinions that might be expressed in the paper do not necessarily represent the views of the U.S. Department of Energy or the United States Government. SAND2026-15952O.

D.I.B. is supported by Army Research Office (ARO) grant W911NF-23-1-0288. The views and conclusions contained in this document are those of the authors and should not be interpreted as representing the official policies, either expressed or implied, of ARO or the U.S. Government. The U.S. Government is authorized to reproduce and distribute reprints for Government purposes notwithstanding any copyright notation herein.

J.M.L is supported by the Army Research Laboratory under Cooperative Agreement Number W911NF-26-2-A015. The views and conclusions contained in this document are those of the authors and should not be interpreted as representing the official policies, either expressed or implied, of the Army Research Laboratory or the U.S. Government. The U.S. Government is authorized to reproduce and distribute reprints for Government purposes notwithstanding any copyright notation herein.

\end{acknowledgments}

\bibliography{main}

@article{lukin_spectral_2024,
	title = {Spectral gaps of two- and three-dimensional many-body quantum systems in the thermodynamic limit},
	volume = {6},
	doi = {10.1103/PhysRevResearch.6.023128},
	number = {2},
	urldate = {2024-05-07},
	journal = {Phys. Rev. Research},
	author = {Lukin, Illya V. and Sotnikov, Andrii G. and Leamer, Jacob M. and Magann, Alicia B. and Bondar, Denys I.},
	month = may,
	year = {2024},
	pages = {023128},
}

@misc{bilokon_dispersion_2025,
      title={Dispersion Relations in Two- and Three-Dimensional Quantum Systems}, 
      author={Valeriia Bilokon and Elvira Bilokon and Illya Lukin and Andrii Sotnikov and Denys Bondar},
      year={2025},
      eprint={2509.15483},
      archivePrefix={arXiv},
      primaryClass={quant-ph},
      url={https://arxiv.org/abs/2509.15483}, 
}

@article{lehtovaara_solution_2007,
	title = {Solution of time-independent {Schrödinger} equation by the imaginary time propagation method},
	volume = {221},
	issn = {00219991},
	url = {https://linkinghub.elsevier.com/retrieve/pii/S0021999106002798},
	doi = {10.1016/j.jcp.2006.06.006},
	number = {1},
	urldate = {2022-06-15},
	journal = {J. Comput. Phys.},
	author = {Lehtovaara, L. and Toivanen, J. and Eloranta, J.},
	month = jan,
	year = {2007},
	pages = {148--157},
}

@article{blume_excited_1997,
	title = {Excited states by quantum {Monte} {Carlo} methods: {Imaginary} time evolution with projection operators},
	volume = {55},
	shorttitle = {Excited states by quantum {Monte} {Carlo} methods},
	url = {https://link.aps.org/doi/10.1103/PhysRevE.55.3664},
	doi = {10.1103/PhysRevE.55.3664},
	number = {3},
	urldate = {2022-06-15},
	journal = {Phys. Rev. E},
	author = {Blume, D. and Lewerenz, M. and Niyaz, P. and Whaley, K. B.},
	month = mar,
	year = {1997},
	pages = {3664--3675},
}

@article{luchow_computing_2003,
	title = {Computing {Energy} {Levels} by {Inversion} of {Imaginary}-{Time} {Cross}-{Correlation} {Functions}},
	volume = {107},
	issn = {1089-5639, 1520-5215},
	url = {https://pubs.acs.org/doi/10.1021/jp034381p},
	doi = {10.1021/jp034381p},
	number = {37},
	urldate = {2022-06-15},
	journal = {J. Phys. Chem. A},
	author = {Lüchow, Arne and Neuhauser, Daniel and Ka, Jaejin and Baer, Roi and Chen, Jianhan and Mandelshtam, Vladimir A.},
	month = sep,
	year = {2003},
	pages = {7175--7180},
}

@article{turro_imaginary-time_2022,
	title = {Imaginary-time propagation on a quantum chip},
	volume = {105},
	issn = {2469-9926, 2469-9934},
	url = {https://link.aps.org/doi/10.1103/PhysRevA.105.022440},
	doi = {10.1103/PhysRevA.105.022440},
	number = {2},
	urldate = {2022-06-29},
	journal = {Phys. Rev. A},
	author = {Turro, F. and Roggero, A. and Amitrano, V. and Luchi, P. and Wendt, K. A. and Dubois, J. L. and Quaglioni, S. and Pederiva, F.},
	month = feb,
	year = {2022},
	pages = {022440},
}

@article{hunt_quantum_2018,
	title = {Quantum {Monte} {Carlo} calculations of energy gaps from first principles},
	volume = {98},
	issn = {2469-9950, 2469-9969},
	url = {https://link.aps.org/doi/10.1103/PhysRevB.98.075122},
	doi = {10.1103/PhysRevB.98.075122},
	number = {7},
	urldate = {2022-07-06},
	journal = {Phys. Rev. B},
	author = {Hunt, R. J. and Szyniszewski, M. and Prayogo, G. I. and Maezono, R. and Drummond, N. D.},
	month = aug,
	year = {2018},
	pages = {075122},
}

@misc{volkin_iterated_nodate,
	title = {Iterated commutators and functions of operators},
	author = {Volkin, H C},
	pages = {19},
        year = {1968},
        howpublished = {\url{https://ntrs.nasa.gov/citations/19680027053}}
}

@article{chin_any_2009,
	title = {Any order imaginary time propagation method for solving the {Schrödinger} equation},
	volume = {470},
	issn = {0009-2614},
	url = {https://www.sciencedirect.com/science/article/pii/S0009261409001080},
	doi = {10.1016/j.cplett.2009.01.068},
	number = {4},
	urldate = {2022-08-02},
	journal = {Chem. Phys. Lett.},
	author = {Chin, Siu A. and Janecek, S. and Krotscheck, E.},
	month = mar,
	year = {2009},
	pages = {342--346},
}

@misc{quspin_git,
    author = {Weinberg, Phillip and Bukov, Marin},
    title = {QuSpin},
    year = {2021},
    howpublished = {\url{https://github.com/weinbe58/QuSpin}}
}

@article{cubitt2015undecidability,
  title={Undecidability of the spectral gap},
  author={Cubitt, Toby S and Perez-Garcia, David and Wolf, Michael M},
  journal={Nature},
  volume={528},
  number={7581},
  pages={207--211},
  year={2015},
  publisher={Nature}
}

@article{balents_spin_2010,
	title = {Spin liquids in frustrated magnets},
	volume = {464},
	copyright = {2010 Nature Publishing Group},
	issn = {1476-4687},
	url = {https://www.nature.com/articles/nature08917},
	doi = {10.1038/nature08917},
	number = {7286},
	urldate = {2023-02-03},
	journal = {Nature},
	author = {Balents, Leon},
	month = mar,
	year = {2010},
	pages = {199--208},
}

@article{han_fractionalized_2012,
	title = {Fractionalized excitations in the spin-liquid state of a kagome-lattice antiferromagnet},
	volume = {492},
	copyright = {2012 Nature Publishing Group, a division of Macmillan Publishers Limited. All Rights Reserved.},
	issn = {1476-4687},
	url = {https://www.nature.com/articles/nature11659},
	doi = {10.1038/nature11659},
	number = {7429},
	urldate = {2023-02-03},
	journal = {Nature},
	author = {Han, Tian-Heng and Helton, Joel S. and Chu, Shaoyan and Nocera, Daniel G. and Rodriguez-Rivera, Jose A. and Broholm, Collin and Lee, Young S.},
	month = dec,
	year = {2012},
	pages = {406--410},
}

@article{motta_determining_2020,
	title = {Determining eigenstates and thermal states on a quantum computer using quantum imaginary time evolution},
	volume = {16},
	issn = {1745-2473, 1745-2481},
	url = {http://www.nature.com/articles/s41567-019-0704-4},
	doi = {10.1038/s41567-019-0704-4},
	number = {2},
	urldate = {2022-08-02},
	journal = {Nat. Phys.},
	author = {Motta, Mario and Sun, Chong and Tan, Adrian T. K. and O’Rourke, Matthew J. and Ye, Erika and Minnich, Austin J. and Brandão, Fernando G. S. L. and Chan, Garnet Kin-Lic},
	month = feb,
	year = {2020},
	pages = {205--210},
}

@misc{Qiskit,
       author = {{Qiskit contributors}},
       title = {Qiskit: An Open-source Framework for Quantum Computing},
       year = {2021},
       doi = {10.5281/zenodo.2573505}
}

@article{jacquemin_excited-state_2011,
	title = {Excited-state calculations with {TD}-{DFT}: from benchmarks to simulations in complex environments},
	volume = {13},
	shorttitle = {Excited-state calculations with {TD}-{DFT}},
	url = {https://pubs.rsc.org/en/content/articlelanding/2011/cp/c1cp22144b},
	doi = {10.1039/C1CP22144B},
	number = {38},
	urldate = {2023-02-08},
	journal = {Phys. Chem. Chem. Phys.},
	author = {Jacquemin, Denis and Mennucci, Benedetta and Adamo, Carlo},
	year = {2011},
	pages = {16987--16998},
}

@article{ramos_low-lying_2018,
	title = {Low-lying excited states by constrained {DFT}},
	volume = {148},
	issn = {0021-9606},
	url = {https://aip.scitation.org/doi/full/10.1063/1.5018615},
	doi = {10.1063/1.5018615},
	number = {14},
	urldate = {2023-02-08},
	journal = {J. Chem. Phys.},
	author = {Ramos, Pablo and Pavanello, Michele},
	month = apr,
	year = {2018},
	pages = {144103},
}

@book{ferre_density-functional_2016,
	address = {Cham},
	series = {Topics in {Current} {Chemistry}},
	title = {Density-{Functional} {Methods} for {Excited} {States}},
	volume = {368},
	isbn = {978-3-319-22080-2 978-3-319-22081-9},
	url = {https://link.springer.com/10.1007/978-3-319-22081-9},
	urldate = {2023-02-08},
	publisher = {Springer International Publishing},
	editor = {Ferré, Nicolas and Filatov, Michael and Huix-Rotllant, Miquel},
	year = {2016},
	doi = {10.1007/978-3-319-22081-9},
}

@article{kaldor_degenerate_1975,
	title = {Degenerate many‐body perturbation theory: {Excited} states of {H} $_{\textrm{2}}$},
	volume = {63},
	issn = {0021-9606, 1089-7690},
	shorttitle = {Degenerate many‐body perturbation theory},
	url = {http://aip.scitation.org/doi/10.1063/1.431600},
	doi = {10.1063/1.431600},
	number = {5},
	urldate = {2023-02-09},
	journal = {J. Chem. Phys.},
	author = {Kaldor, Uzi},
	month = sep,
	year = {1975},
	pages = {2199--2205},
}

@article{hybertsen_first-principles_1985,
	title = {First-{Principles} {Theory} of {Quasiparticles}: {Calculation} of {Band} {Gaps} in {Semiconductors} and {Insulators}},
	volume = {55},
	shorttitle = {First-{Principles} {Theory} of {Quasiparticles}},
	url = {https://link.aps.org/doi/10.1103/PhysRevLett.55.1418},
	doi = {10.1103/PhysRevLett.55.1418},
	number = {13},
	urldate = {2023-02-09},
	journal = {Phys. Rev. Lett.},
	author = {Hybertsen, Mark S. and Louie, Steven G.},
	month = sep,
	year = {1985},
	pages = {1418--1421},
}

@article{hybertsen_electron_1986,
	title = {Electron correlation in semiconductors and insulators: {Band} gaps and quasiparticle energies},
	volume = {34},
	shorttitle = {Electron correlation in semiconductors and insulators},
	url = {https://link.aps.org/doi/10.1103/PhysRevB.34.5390},
	doi = {10.1103/PhysRevB.34.5390},
	number = {8},
	urldate = {2023-02-09},
	journal = {Phys. Rev. B.},
	author = {Hybertsen, Mark S. and Louie, Steven G.},
	month = oct,
	year = {1986},
	pages = {5390--5413},
}

@article{godby_accurate_1986,
	title = {Accurate {Exchange}-{Correlation} {Potential} for {Silicon} and {Its} {Discontinuity} on {Addition} of an {Electron}},
	volume = {56},
	url = {https://link.aps.org/doi/10.1103/PhysRevLett.56.2415},
	doi = {10.1103/PhysRevLett.56.2415},
	number = {22},
	urldate = {2023-02-09},
	journal = {Phys. Rev. Lett.},
	author = {Godby, R. W. and Schlüter, M. and Sham, L. J.},
	month = jun,
	year = {1986},
	pages = {2415--2418},
}

@article{godby_self-energy_1988,
	title = {Self-energy operators and exchange-correlation potentials in semiconductors},
	volume = {37},
	url = {https://link.aps.org/doi/10.1103/PhysRevB.37.10159},
	doi = {10.1103/PhysRevB.37.10159},
	number = {17},
	urldate = {2023-02-09},
	journal = {Phys. Rev. B},
	author = {Godby, R. W. and Schlüter, M. and Sham, L. J.},
	month = jun,
	year = {1988},
	pages = {10159--10175},
}

@article{hartmann_excitation_2006,
	title = {Excitation and entanglement transfer versus spectral gap},
	volume = {8},
	issn = {1367-2630},
	url = {https://iopscience.iop.org/article/10.1088/1367-2630/8/6/094},
	doi = {10.1088/1367-2630/8/6/094},
	number = {6},
	urldate = {2023-02-09},
	journal = {New J. Phys.},
	author = {Hartmann, M J and Reuter, M E and Plenio, M B},
	month = jun,
	year = {2006},
	pages = {94--94},
}

@article{yang_topological_2013,
	title = {Topological phase transitions with and without energy gap closing},
	volume = {15},
	issn = {1367-2630},
	url = {https://dx.doi.org/10.1088/1367-2630/15/8/083042},
	doi = {10.1088/1367-2630/15/8/083042},
	number = {8},
	urldate = {2023-02-09},
	journal = {New J. Phys.},
	author = {Yang, Yunyou and Li, Huichao and Sheng, L. and Shen, R. and Sheng, D. N. and Xing, D. Y.},
	month = aug,
	year = {2013},
	pages = {083042},
}

@article{hastings_spectral_2006,
	title = {Spectral {Gap} and {Exponential} {Decay} of {Correlations}},
	volume = {265},
	issn = {1432-0916},
	url = {https://doi.org/10.1007/s00220-006-0030-4},
	doi = {10.1007/s00220-006-0030-4},
	number = {3},
	urldate = {2023-02-09},
	journal = {Commun. Math. Phys.},
	author = {Hastings, Matthew B. and Koma, Tohru},
	month = aug,
	year = {2006},
	pages = {781--804},
}

@article{dziarmaga_dynamics_2005,
	title = {Dynamics of a {Quantum} {Phase} {Transition}: {Exact} {Solution} of the {Quantum} {Ising} {Model}},
	volume = {95},
	issn = {0031-9007, 1079-7114},
	shorttitle = {Dynamics of a {Quantum} {Phase} {Transition}},
	url = {https://link.aps.org/doi/10.1103/PhysRevLett.95.245701},
	doi = {10.1103/PhysRevLett.95.245701},
	number = {24},
	urldate = {2023-02-10},
	journal = {Phys. Rev. Lett.},
	author = {Dziarmaga, Jacek},
	month = dec,
	year = {2005},
	pages = {245701},
}

@book{sachdev_quantum_1999,
	address = {Cambridge ; New York},
	title = {Quantum phase transitions},
	isbn = {978-0-521-58254-4},
	publisher = {Cambridge University Press},
	author = {Sachdev, Subir},
	year = {1999},
}

@article{kopec_instabilities_1989,
	title = {Instabilities in the quantum {Sherrington}-{Kirkpatrick} {Ising} spin glass in transverse and longitudinal fields},
	volume = {39},
	url = {https://link.aps.org/doi/10.1103/PhysRevB.39.12418},
	doi = {10.1103/PhysRevB.39.12418},
	number = {16},
	urldate = {2023-02-10},
	journal = {Phys. Rev. B},
	author = {Kopeć, T. K. and Usadel, K. D. and Büttner, G.},
	month = jun,
	year = {1989},
	pages = {12418--12421},
}

@article{laumann_cavity_2008,
	title = {Cavity method for quantum spin glasses on the {Bethe} lattice},
	volume = {78},
	url = {https://link.aps.org/doi/10.1103/PhysRevB.78.134424},
	doi = {10.1103/PhysRevB.78.134424},
	number = {13},
	urldate = {2023-02-10},
	journal = {Phys. Rev. B},
	author = {Laumann, C. and Scardicchio, A. and Sondhi, S. L.},
	month = oct,
	year = {2008},
	pages = {134424},
}

@misc{farhi_quantum_2000,
      title={Quantum Computation by Adiabatic Evolution}, 
      author={Edward Farhi and Jeffrey Goldstone and Sam Gutmann and Michael Sipser},
      year={2000},
      eprint={quant-ph/0001106},
      archivePrefix={arXiv},
      primaryClass={quant-ph},
      url={https://arxiv.org/abs/quant-ph/0001106}, 
}

@article{boixo_evidence_2014,
	title = {Evidence for quantum annealing with more than one hundred qubits},
	volume = {10},
	copyright = {2014 Nature Publishing Group},
	issn = {1745-2481},
	url = {https://www.nature.com/articles/nphys2900},
	doi = {10.1038/nphys2900},
	number = {3},
	urldate = {2023-02-10},
	journal = {Nat. Phys.},
	author = {Boixo, Sergio and Rønnow, Troels F. and Isakov, Sergei V. and Wang, Zhihui and Wecker, David and Lidar, Daniel A. and Martinis, John M. and Troyer, Matthias},
	month = mar,
	year = {2014},
	pages = {218--224},
}

@article{troels_2014,
author = {Troels F. Rønnow  and Zhihui Wang  and Joshua Job  and Sergio Boixo  and Sergei V. Isakov  and David Wecker  and John M. Martinis  and Daniel A. Lidar  and Matthias Troyer },
title = {Defining and detecting quantum speedup},
journal = {Science},
volume = {345},
number = {6195},
pages = {420-424},
year = {2014},
doi = {10.1126/science.1252319},
URL = {https://www.science.org/doi/abs/10.1126/science.1252319}}

@misc{shin_how_2014,
      title={How ``Quantum" is the D-Wave Machine?}, 
      author={Seung Woo Shin and Graeme Smith and John A. Smolin and Umesh Vazirani},
      year={2014},
      eprint={1401.7087},
      archivePrefix={arXiv},
      primaryClass={quant-ph},
      url={https://arxiv.org/abs/1401.7087}, 
}

@article{anderson_1987,
author = {P. W. Anderson },
title = {The Resonating Valence Bond State in {La}$_2${Cu}{O}$_4$ and Superconductivity},
journal = {Science},
volume = {235},
number = {4793},
pages = {1196-1198},
year = {1987},
doi = {10.1126/science.235.4793.1196},
}

@article{imada_1998,
  title = {Metal-insulator transitions},
  author = {Imada, Masatoshi and Fujimori, Atsushi and Tokura, Yoshinori},
  journal = {Rev. Mod. Phys.},
  volume = {70},
  issue = {4},
  pages = {1039--1263},
  numpages = {0},
  year = {1998},
  month = {Oct},
  publisher = {American Physical Society},
  doi = {10.1103/RevModPhys.70.1039},
  url = {https://link.aps.org/doi/10.1103/RevModPhys.70.1039}
}

@article{hofrichter_2016,
  title = {Direct Probing of the Mott Crossover in the $\mathrm{SU}(N)$ Fermi-Hubbard Model},
  author = {Hofrichter, Christian and Riegger, Luis and Scazza, Francesco and H\"ofer, Moritz and Fernandes, Diogo Rio and Bloch, Immanuel and F\"olling, Simon},
  journal = {Phys. Rev. X},
  volume = {6},
  issue = {2},
  pages = {021030},
  numpages = {8},
  year = {2016},
  month = {Jun},
  publisher = {American Physical Society},
  doi = {10.1103/PhysRevX.6.021030},
  url = {https://link.aps.org/doi/10.1103/PhysRevX.6.021030}
}

@misc{leamer_git,
    author = {Jacob M. Leamer},
    title = {Spectral Gaps},
    year = {2020},
    howpublished = {\url{https://github.com/jleamer/spectral_gaps}}
}

@misc{leamer_quantum_2024,
      title={Quantum Dynamical Emulation of Imaginary Time Evolution}, 
      author={Jacob M. Leamer and Alicia B. Magann and Denys I. Bondar and Gerard McCaul},
      year={2024},
      eprint={2403.03350},
      archivePrefix={arXiv},
      primaryClass={quant-ph},
      url={https://arxiv.org/abs/2403.03350}, 
}

@article{kamakari_digital_2022,
	title = {Digital {Quantum} {Simulation} of {Open} {Quantum} {Systems} {Using} {Quantum} {Imaginary}--{Time} {Evolution}},
	volume = {3},
	url = {https://link.aps.org/doi/10.1103/PRXQuantum.3.010320},
	doi = {10.1103/PRXQuantum.3.010320},
	number = {1},
	urldate = {2024-02-16},
	journal = {PRX Quantum},
	author = {Kamakari, Hirsh and Sun, Shi-Ning and Motta, Mario and Minnich, Austin J.},
	month = feb,
	year = {2022},
	pages = {010320},
}

@article{cleve_quantum_1998,
	title = {Quantum algorithms revisited},
	volume = {454},
	url = {https://royalsocietypublishing.org/doi/10.1098/rspa.1998.0164},
	doi = {10.1098/rspa.1998.0164},
	number = {1969},
	urldate = {2024-01-05},
	journal = {Proceedings of the Royal Society of London. Series A: Mathematical, Physical and Engineering Sciences},
	author = {Cleve, R. and Ekert, A. and Macchiavello, C. and Mosca, M.},
	month = {Jan},
	year = {1998},
	pages = {339--354},
}

@misc{McCaul2025FreeSnacks,
      title={Free Snacks in Quantum Complexity}, 
      author={Gerard McCaul},
      year={2025},
      eprint={2509.04618},
      archivePrefix={arXiv},
      primaryClass={quant-ph},
      url={https://arxiv.org/abs/2509.04618}, 
}

@misc{Wu2025EffectiveHamiltonians,
      title={Quantum algorithm for solving nonlinear differential equations based on physics-informed effective Hamiltonians}, 
      author={Hsin-Yu Wu and Annie E. Paine and Evan Philip and Antonio A. Gentile and Oleksandr Kyriienko},
      year={2025},
      eprint={2504.13174},
      archivePrefix={arXiv},
      primaryClass={quant-ph},
      url={https://arxiv.org/abs/2504.13174}, 
}

@misc{McMahon2025QITEFlows,
      title={Equating quantum imaginary time evolution, Riemannian gradient flows, and stochastic implementations}, 
      author={Nathan A. McMahon and Mahum Pervez and Christian Arenz},
      year={2025},
      eprint={2504.06123},
      archivePrefix={arXiv},
      primaryClass={quant-ph},
      url={https://arxiv.org/abs/2504.06123}, 
}

@misc{Alipour2025StateBasedITE,
      title={State-Based Quantum Simulation of Imaginary-Time Evolution}, 
      author={S. Alipour and T. Ojanen},
      year={2025},
      eprint={2506.12381},
      archivePrefix={arXiv},
      primaryClass={quant-ph},
      url={https://arxiv.org/abs/2506.12381}, 
}

@article{Joseph2020KvN,
  title = {Koopman--von Neumann approach to quantum simulation of nonlinear classical dynamics},
  author = {Joseph, Ilon},
  journal = {Phys. Rev. Res.},
  volume = {2},
  issue = {4},
  pages = {043102},
  numpages = {17},
  year = {2020},
  month = {Oct},
  publisher = {American Physical Society},
  doi = {10.1103/PhysRevResearch.2.043102},
  url = {https://link.aps.org/doi/10.1103/PhysRevResearch.2.043102}
}

@misc{NovikauJoseph2025AdvecDiff,
      title={An efficient explicit implementation of a near-optimal quantum algorithm for simulating linear dissipative differential equations}, 
      author={Ivan Novikau and Ilon Joseph},
      year={2025},
      eprint={2501.11146},
      archivePrefix={arXiv},
      primaryClass={quant-ph},
      url={https://arxiv.org/abs/2501.11146}, 
}

@misc{NovikauJoseph2025Carleman,
      title={Globalizing the Carleman linear embedding method for nonlinear dynamics}, 
      author={Ivan Novikau and Ilon Joseph},
      year={2025},
      eprint={2510.15715},
      archivePrefix={arXiv},
      primaryClass={quant-ph},
      url={https://arxiv.org/abs/2510.15715}, 
}

@article{Joseph2023Fusion,
    author = {Joseph, I. and Shi, Y. and Porter, M. D. and Castelli, A. R. and Geyko, V. I. and Graziani, F. R. and Libby, S. B. and DuBois, J. L.},
    title = {Quantum computing for fusion energy science applications},
    journal = {Physics of Plasmas},
    volume = {30},
    number = {1},
    pages = {010501},
    year = {2023},
    month = {01},
    issn = {1070-664X},
    doi = {10.1063/5.0123765},
    url = {https://doi.org/10.1063/5.0123765},
}

@article{ParkerJoseph2020GenEigQPE,
  title = {Quantum phase estimation for a class of generalized eigenvalue problems},
  author = {Parker, Jeffrey B. and Joseph, Ilon},
  journal = {Phys. Rev. A},
  volume = {102},
  issue = {2},
  pages = {022422},
  numpages = {5},
  year = {2020},
  month = {Aug},
  publisher = {American Physical Society},
  doi = {10.1103/PhysRevA.102.022422},
  url = {https://link.aps.org/doi/10.1103/PhysRevA.102.022422}
}

@article{Kempe2006LocalHam,
  author       = {Kempe, Julia and Kitaev, Alexei and Regev, Oded},
  title        = {The Complexity of the Local Hamiltonian Problem},
  journal      = {SIAM Journal on Computing},
  year         = {2006},
  volume       = {35},
  number       = {5},
  pages        = {1070--1097},
  doi          = {10.1137/S0097539704445226},
}

@article{Troyer2005SignProblem,
  author       = {Troyer, Matthias and Wiese, Uwe{-}Jens},
  title        = {Computational Complexity and Fundamental Limitations to Fermionic Quantum Monte Carlo Simulations},
  journal      = {Physical Review Letters},
  year         = {2005},
  volume       = {94},
  number       = {17},
  pages        = {170201},
  doi          = {10.1103/PhysRevLett.94.170201},
}

@article{Orus2014TN,
  author       = {Or{\'u}s, Rom{\'a}n},
  title        = {A Practical Introduction to Tensor Networks: Matrix Product States and Projected Entangled Pair States},
  journal      = {Annals of Physics},
  year         = {2014},
  volume       = {349},
  pages        = {117--158},
  doi          = {10.1016/j.aop.2014.06.013},
}

@article{McCaul2020DrivenImposters,
  title   = {Driven Imposters: Controlling Expectations in Many-Body Systems},
  author  = {McCaul, Gerard and Orthodoxou, Christopher and Jacobs, Kurt and Booth, George H. and Bondar, Denys I.},
  journal = {Phys. Rev. Lett.},
  volume  = {124},
  number  = {18},
  pages   = {183201},
  year    = {2020},
  doi     = {10.1103/PhysRevLett.124.183201},
  eprint  = {2010.05399},
  archivePrefix = {arXiv}
}

@article{McCaul2020ControllingArbitrary,
  title   = {Controlling Arbitrary Observables in Correlated Many-Body Systems},
  author  = {McCaul, Gerard and Orthodoxou, Christopher and Jacobs, Kurt and Booth, George H. and Bondar, Denys I.},
  journal = {Phys. Rev. A},
  volume  = {101},
  number  = {5},
  pages   = {053408},
  year    = {2020},
  doi     = {10.1103/PhysRevA.101.053408}
}

@book{NegeleOrland,
  author    = {J. W. Negele and H. Orland},
  title     = {Quantum Many-Particle Systems},
  publisher = {Westview Press},
  year      = {1998}
}

@book{AltlandSimons,
  author    = {Alexander Altland and Ben Simons},
  title     = {Condensed Matter Field Theory},
  publisher = {Cambridge University Press},
  edition   = {2},
  year      = {2010}
}

@book{feynman_hibbs_1965,
  author    = {R. P. Feynman and A. R. Hibbs},
  title     = {Quantum Mechanics and Path Integrals},
  year      = {1965},
  publisher = {McGraw-Hill},
}

@book{negele_orland_1988,
  author    = {J. W. Negele and H. Orland},
  title     = {Quantum Many-Particle Systems},
  year      = {1988},
  publisher = {Addison-Wesley},
}

@article{ceperley_1995,
  title = {Path integrals in the theory of condensed helium},
  author = {Ceperley, D. M.},
  journal = {Rev. Mod. Phys.},
  volume = {67},
  issue = {2},
  pages = {279--355},
  numpages = {0},
  year = {1995},
  month = {Apr},
  publisher = {American Physical Society},
  doi = {10.1103/RevModPhys.67.279},
  url = {https://link.aps.org/doi/10.1103/RevModPhys.67.279}
}

@article{foulkes_2001,
  title = {Quantum Monte Carlo simulations of solids},
  author = {Foulkes, W. M. C. and Mitas, L. and Needs, R. J. and Rajagopal, G.},
  journal = {Rev. Mod. Phys.},
  volume = {73},
  issue = {1},
  pages = {33--83},
  numpages = {0},
  year = {2001},
  month = {Jan},
  publisher = {American Physical Society},
  doi = {10.1103/RevModPhys.73.33},
  url = {https://link.aps.org/doi/10.1103/RevModPhys.73.33}
}

@article{mccaul_how_2021,
	title = {How to win friends and influence functionals: deducing stochasticity from deterministic dynamics},
	volume = {230},
	issn = {1951-6355, 1951-6401},
	shorttitle = {How to win friends and influence functionals},
	url = {https://link.springer.com/10.1140/epjs/s11734-021-00068-2},
	doi = {10.1140/epjs/s11734-021-00068-2},
	number = {4},
	urldate = {2024-07-09},
	journal = {The European Physical Journal Special Topics},
	author = {McCaul, Gerard and Bondar, Denys. I.},
	month = jun,
	year = {2021},
	pages = {733--754}
}

@article{mccaul_partition-free_2017,
	title = {Partition-free approach to open quantum systems in harmonic environments: {An} exact stochastic {Liouville} equation},
	volume = {95},
	copyright = {http://link.aps.org/licenses/aps-default-license},
	issn = {2469-9950, 2469-9969},
	url = {https://link.aps.org/doi/10.1103/PhysRevB.95.125124},
	doi = {10.1103/PhysRevB.95.125124},
	number = {12},
	urldate = {2024-07-09},
	journal = {Physical Review B},
	author = {McCaul, G. M. G. and Lorenz, C. D. and Kantorovich, L.},
	month = mar,
	year = {2017},
	pages = {125124}

}

@article {Williams371,
	author = {Williams, R. T. and Bridges, J. W.},
	title = {Fluorescence of solutions: A review},
	volume = {17},
	number = {4},
	pages = {371--394},
	year = {1964},
	doi = {10.1136/jcp.17.4.371},
	publisher = {BMJ Publishing Group},
	issn = {0021-9746},
	URL = {https://jcp.bmj.com/content/17/4/371},
	journal = {Journal of Clinical Pathology}
}

\appendix
\section{Commutator Expansion \label{app:comm}}
We wish to demonstrate via induction that the $M$-th nested commutator of two general operators $H$ and $O$ is given by \cite{volkin_iterated_nodate} 
\begin{align}
    [H, O]_M = \sum_{m=0}^M(-1)^m\binom{M}{m}H^{M-m}O H^m, \label{eq:comm_exp}
\end{align}
for any whole number $M$.  Consider $[H, O] = HO - OH$.  This can also be written as
\begin{align}
    [H, O] &= \sum_{m=0}^{M=1}(-1)^m\binom{M}{m}H^{1-m}O H^m, \\
                &= HO - O H.
\end{align}
Suppose this is true for the $M$-th commutator.  Now consider calculating the commutator of $H$ with the previous expression to obtain the $M+1$-th commutator.  Then we would have
\begin{align}
    [H, O]_{M+1} &= \sum_{m=0}^M(-1)^m\binom{M}{m}H^{M-m+1}O H^m +  \notag \\ &\sum_{m'=0}^M(-1)^{m'+1}\binom{M}{m'}H^{M-m'}O H^{m'+1}. \label{eq:total_expr}
\end{align}
Terms where $m=m'+1$ can be simplified, but this requires separating the $m=0$ and $m'=M$ terms from the sums to obtain
\begin{align}
    [H, O]_{M+1} &= H^{M+1}O + X + (-1)^{M+1}O H^{M+1},
\end{align}
where $X$ is the summation containing like terms,
\begin{align}
    X &= \sum_{m'=0}^{M-1}(-1)^{m'+1}H^{M-m'}O H^{m'+1}\notag \\ &\left[\binom{M}{m'+1} + \binom{M}{m'}\right].
\end{align}
Taking advantage of the fact that $\binom{n}{k} = \binom{n-1}{k} + \binom{n-1}{k-1}$ and adjusting indices to be clearer yields
\begin{align}
    X &= \sum_{k=1}^M(-1)^k\binom{M+1}{k}H^{M+1-k}O H^k,
\end{align}
Combining everything in Eq.~\eqref{eq:total_expr}, we have
\begin{align}
    [H, O]_{M+1} &= H^{M+1}O \notag \\ &+\sum_{k=1}^M(-1)^k\binom{M+1}{k}H^{M+1-k}O H^k \notag \\ &+(-1)^{M+1}O H^{M+1},
\end{align}
or
\begin{align}
    [H, O]_{M+1} &= \sum_{k=0}^{M+1}(-1)^k\binom{M+1}{k}H^{M+1-k}O H^k.
\end{align}
Thus, by induction, the expansion of the $M$-th order commutator is given by Eq.~\eqref{eq:comm_exp} for any $M$.

\end{document}